\documentclass[aps,nofootinbib,preprint]{revtex4}

\usepackage{graphicx}
\usepackage{graphics}
\usepackage{amsmath}
\usepackage{subfigure}

\begin{document}
\title{Constraining the time variation of the coupling constants from cosmic microwave background: effect of $\Lambda_{\mathrm{QCD}}$}

\author{Masahiro Nakashima $^{1,2}$}
\author{Kazuhide Ichikawa $^{3}$}
\author{Ryo Nagata $^{2}$}
\author{Jun'ichi Yokoyama $^{2,4}$}

\affiliation{$^{1}$ Department of Physics, Graduate School of Science,\\  
The University of Tokyo, Tokyo 113-0033, Japan \\
$^{2}$ Research Center for the Early Universe (RESCEU), \\
Graduate School of Science, The University of Tokyo, Tokyo 113-0033, Japan \\
$^{3}$ Department of Micro Engineering, Kyoto University, Kyoto 606-8501, Japan \\
$^{4}$ Institute for the Physics and Mathematics of the Universe(IPMU), \\
The University of Tokyo, Kashiwa, Chiba, 277-8568, Japan}

\begin{abstract}
We investigate constraints on the time variation of the fine structure constant between the recombination epoch and the present epoch, $\Delta\alpha/\alpha \equiv (\alpha_{\mathrm{rec}} - \alpha_{\mathrm{now}})/\alpha_{\mathrm{now}}$, from cosmic microwave background (CMB) taking into account simultaneous variation of other physical constants, namely the electron mass $m_{e}$ and the proton mass $m_{p}$. In other words, we consider the variation of Yukawa coupling and the QCD scale $\Lambda_{\mathrm{QCD}}$ in addition to the electromagnetic coupling. We clarify which parameters can be determined from CMB temperature anisotropy in terms of singular value decomposition. Assuming a relation among variations of coupling constants governed by a single scalar field (the dilaton), the 95\,\% confidence level (C.L.) constraint on $\Delta\alpha/\alpha$ is found to be $-8.28 \times 10^{-3} < \Delta\alpha/\alpha < 1.81 \times 10^{-3}$, which is tighter than the one obtained by considering only the change of $\alpha$ and $m_{e}$. We also obtain the constraint on the time variation of the proton-to-electron mass ratio $\mu \equiv m_{p}/m_{e}$ to be $-0.52 < \Delta\mu/\mu < 0.17$ (95\,\% C.L.) under the same assumption. Finally, we also implement a forecast for constraints from the PLANCK survey.
\end{abstract}

\begin{flushright}
RESCEU-25-09
\end{flushright}

\maketitle

\section{Introduction}
 Whether or not the physical constants are truly constant is not only a fundamental issue in physics but also an important probe of the theories of extra dimensions. Since the early studies of Dirac \cite{Dirac:1937ti,Dirac:1938mt}, many theoretical models that could accomodate the variation of physical constants have been proposed \cite{Bekenstein:1982eu,Barrow:2001iw,Barrow:2005qf,Uzan:2002vq, Chiba:2006xx}. Unification theories such as superstring theories predict the existence of an additional scalar field $\Phi$ called ``dilaton" and the destabilization of this field \cite{Dine:1985he,Brustein:1992nk} might cause the variation of coupling constants. In more phenomenological contexts, dynamical scalar fields introduced to explain the recent discovery of the accerelated expansion may be non-minimally coupled to gauge fields in the Standard model, which may also lead to time varying coupling constants \cite{Avelino:2006gc,Avelino:2008dc,Fujii:2007zg,Fujii:2008zj,Dent:2008vd,Avelino:2009fd}. Thus, investigating the variation of the coupling constants can be an important probe of these theories.

 Not only these theoretical sides but also the observational analysis of various systems have been playing an important role in the investigation of the time variation of physical constants. Among them, the claim for the deviation of the past value of the fine structure constant $\alpha$ from the present one by using the high-redshift quasar absorption system \cite{Webb:1998cq,Webb:2000mn} had the great impact on this topic. Further analysis for the quasar absorption spectra gives the support for this positive result of a cosmological variation of $\alpha$ \cite{Murphy:2004,Murphy:2006vs,Murphy:2007qs,Molaro:2007kp}, though some authors takes objection to such variations \cite{Chand:2004ct,Srianand:2004mq,Molaro:2007kp}.

Stimulated by these studies from both theoretical and observational sides, various constraints on the time variation of the fine structure constant $\alpha$ have been investigated from diverse observations. We briefly summarize those terrestrial and celestial limits on $\alpha$ as follows (see \cite{Uzan:2002vq,Chiba:2001ui} for a review). The atomic clocks constrain the current value of the temporal derivative of $\alpha$ as $\dot{\alpha}/\alpha = (-3.3\pm 3.0)\times 10^{-16} \ \mathrm{yr}^{-1}$\cite{Blatt:2008su,Fortier:2007jf,Peik:2004qn,Fischer:2004jt}. The measurement of the frequency ratio of aluminium and mercury single-ion optical clocks  provides $\dot{\alpha}/\alpha=(-1.6\pm 2.3) \times 10^{-17} \ \mathrm{yr}^{-1}$ \cite{Rosenband:2008}. From the analysis of Sm isotopes in the Oklo natural reactor in Gabon, we get two bounds on the variation of $\alpha$ as $\Delta\alpha/\alpha = -(0.8\pm 1.0) \times 10^{-8}$ and $\Delta\alpha/\alpha=(0.88 \pm 0.07) \times 10^{-7}$ \cite{Fujii:2002hc}, which measures the value at the redshift $z\sim 0.1$ constraint. From the spectra of quasars, several limits have been obtained using various data sets: $\Delta\alpha/\alpha = (-0.57\pm 0.11) \times 10^{-5} \ (z \sim 0.2-4.2$) from the Keck/HIRES instrument \cite{Murphy:2004}, $\Delta\alpha/\alpha = (-0.64 \pm 0.36)\times 10^{-5} \ (z\sim 0.4-2.3)$ from the Ultraviolet and Visual Echelle Spectrograph (UVES) instrument \cite{Murphy:2006vs,Murphy:2007qs}. Big bang nucleosynthesis (BBN) provides constraints at very high redshifts ($z\sim 10^{9}-10^{10}$), for example, $-5.0\times 10^{-2} < \Delta\alpha/\alpha < 1.0 \times 10^{-2} \ (95 \% \mathrm{C.L.})$ \cite{Ichikawa:2002bt}. Finally, Cosmic Microwave Background (CMB) measures $\alpha$ at $z\sim 10^{3}$ and the limits from WMAP 1-year, 3-year and 5-year data read $-0.06 < \Delta\alpha/\alpha < 0.01$ \cite{Rocha:2003gc}, $-0.039 < \Delta\alpha/\alpha < 0.010$ \cite{Stefanescu:2007aa} and $-0.028 < \Delta\alpha/\alpha 
 < 0.026$ \cite{Nakashima:2008cb} respectively, all of which are at $95\% \mathrm{C.L.}$ Recently, Menegoni et al. \cite{Menegoni:2009rg} obtained more improved result $-0.013 < \Delta\alpha/\alpha < 0.015$ at $95\% \mathrm{C.L.}$ from WMAP 5-year date combined with ACBAR, QUAD and BICEP experiments data.

 Along with $\alpha$, several observational constraints have been obtained on another physical constant, the proton-to-electron mass ratio $\mu\equiv m_{p}/m_{e}$. As in $\alpha$, atomic clocks very tightly constrain the current value of its temporal derivative as $\dot{\mu}/\mu = (-1.5\pm 1.7)\times 10^{-15} \ \mathrm{yr}^{-1}$\cite{Blatt:2008su}. Reinhold et al. \cite{Reinhold:2006zn} reports the non-vanishing variation of $\mu$ as $\Delta\mu/\mu = (2.4\pm 0.6)\times 10^{-5}$ from a weighted sum of accurate $\mathrm{H}_{2}$ spectral lines in two quasar systems (corresponding redshifts are $z=2.59$ and $z=3.02$). Based on the quasar absorption spectra of $\mathrm{NH}_{3}$, the limit is $\Delta\mu/\mu = (0.6\pm 1.9)\times 10^{-6}$ at $z=0.6847$ \cite{Flambaum:2007fa} and further detailed measurement by the same method tighten the bound as $\Delta\mu/\mu = (0.74 \pm 0.47)\times 10^{-6}$ \cite{Murphy:2008yy}. 
    
 In this paper, using WMAP 5-year data, we focus on the CMB constraint on $\alpha$ in the case that the multiple physical constants may vary with time simultaneously according to the expectation values of a single dilaton field $\Phi$. Although similar analyses has been done for BBN \cite{Ichikawa:2002bt,Campbell:1994bf,Langacker:2001td,Dent:2001ga,Muller:2004gu,Dent:2008gx}, quasar absorption system \cite{Calmet:2001nu,Calmet:2002ja} or the Oklo reactor \cite{Olive:2002tz}, few attempts from CMB data have been made except for several preceding literatures \cite{Ichikawa:2006nm,Landau:2008re,Scoccola:2008jw}, which treated simultaneous variations of only $\alpha$ and $m_{e}$. In addition to $\alpha$ and $m_{e}$, many theoretical models predict the time variation of the QCD scale $\Lambda_{\mathrm{QCD}}$ (which results from the variation of the strong coupling constant), so it is important to include the $\Lambda_{\mathrm{QCD}}$ effect in the analysis. We consider the effect of varying $\Lambda_{\mathrm{QCD}}$ as the change in the proton mass $m_{p}$ which is assumed to be proportional to $\Lambda_{\mathrm{QCD}}$
\footnote{When we consider the time variation of $\Lambda_{\mathrm{QCD}}$, the neutron mass also changes. In this paper, we neglect the difference between the proton mass and the neutron mass, and when we refer to the time variation of the proton mass $m_{p}$, we implicitly include the time variation of the neutron mass assuming that the neutron mass changes in exactly the same way as the proton mass.}
. Thus, we consider simultaneous variation of $\alpha$, $m_{e}$ and $m_{p}$. Since how the variation of $\Lambda_{\mathrm{QCD}}$ is related to that of $\alpha$ or $m_{e}$ depends on the details of the unification model \cite{Dine:2002ir}, we here adopt a specific example of the relation among those physical constants following \cite{Ichikawa:2002bt} or \cite{Campbell:1994bf}, which is based on a low energy effective theory of string theory. The newly incorporated change of $\Lambda_{\mathrm{QCD}}$ leads to a new and tighter constraint on the time variation of $\alpha$. We also obtain a constraint on $\mu$ induced by the time dependence of $\Lambda_{QCD}$ and $m_{e}$. 

 This paper is organized as follows. Section \ref{sec:effect} provides how the variation of each physical constant affects the CMB power spectrum both qualitatively and quantitatively with particular attention to $\Lambda_{\mathrm{QCD}}$. Section \ref{sec:dilaton} briefly describes the model we adopt based on string theory which governs the time evolution of the three physical constants we consider, namely, the variation of the fine structure constant, the electron mass and the proton mass. In section \ref{sec:constraint}, we provide the limits of those variations. Section \ref{sec:conclusion} is devoted to conclusion. The details of the calculations of the degeneracy and parameter dependence are given in the appendix A. In appendix B, we present a forecast for constraints from the PLANCK experiment.

\section{The effect of variations of physical constants on the CMB} \label{sec:effect}
 As is well known, a primary CMB photon is a relic signal from the last scattering surfuce about thirteen billion years ago, which means that the most important process for CMB photon is the recombination. It has been known that changing the value of the fine structure constant $\alpha$ and the electron mass $m_{e}$ affects the CMB power spectrum mainly through the change of the epoch of recombination \cite{Hannestad:1998xp,Kaplinghat:1998ry,Kujat:1999rk,Yoo:2002vw}. The larger value of $\alpha$ or $m_{e}$ at the recombination epoch causes the higher redshift of the last scattering surface 
 because the hydrogen binding energy scales as $\alpha^{2}m_{e}$. This results in three characteristic signatures in the angular power spectrum of the temperature anisotropy, namely, shift of the peaks to higher multipoles, increase of the height of the peaks due to the enhanced early integrated Sachs Wolfe effect, and decrease of the small-scale diffusion damping effect. To be more precise, however, the process of the recombination depend on $\alpha$ and $m_{e}$ in a subtly different manner as will be explicitly shown in eqs.(\ref{4v-1})-(\ref{4v-4}) in the argument of the four CMB-characteristic variables below. 

The time variation of the proton mass $m_{p}$ affects the CMB power spectrum mainly through the baryon density $\rho_B$. In the standard cosmology with no variation of physical constants, the baryon density scales as $\rho_{B,\mathrm{rec}} = \rho_{B,\mathrm{0}} (a_{\mathrm{0}}/a_{\mathrm{rec}})^{3}$, where $a$ is the scale factor and the subscripts ``rec" and ``0" denote values at the epoch of recombination and the present epoch, respectively. When we allow the time variation of $m_p$, there is an additional change as
\begin{equation}
\rho_{B,\mathrm{rec}} = \rho_{B,\mathrm{0}}\left(\frac{a_{\mathrm{0}}}{a_{\mathrm{rec}}}\right)^{3} \frac{m_{p,\mathrm{rec}}}{m_{p,\mathrm{0}}} \equiv \rho_{B,\mathrm{0}}\left(\frac{a_{\mathrm{0}}}{a_{\mathrm{rec}}}\right)^{3} \left(1+\frac{\Delta m_{p}}{m_{p}}\right), \label{omegab_mp}
\end{equation}
where we define $\Delta m_{p} \equiv m_{p,\mathrm{rec}}-m_{p,\mathrm{0}}$. Therefore, the effect of changing $m_{p,\mathrm{rec}}$ is very similar to changing $\rho_{B,\mathrm{0}}$. One may even consider that $m_{p,\mathrm{rec}}$ and $\rho_{B,\mathrm{0}}$ are totally degenerate. However, there appears a subtle difference as regards the baryon number density $n_B$ which affects the recombination process. This is because, contrary to the case of $\rho_B$, the scaling $n_{B,\mathrm{rec}}= n_{B,\mathrm{0}}(a_{\mathrm{0}}/a_{\mathrm{rec}})^{3}$ is not affected by the variation of $m_p$ (note that $n_{B} \equiv \rho_{B}/m_{p}$). Thus, suppose one increases $\rho_{B,\mathrm{0}}$ and decreases $m_{p,\mathrm{rec}}$ while preserving the value of $\rho_{B,\mathrm{rec}}$, $n_{B,\mathrm{rec}}$ is increased.
This is the reason why varying $\rho_{B,\mathrm{0}}$ (as is done in the usual parameter search) and varying $m_{p,\mathrm{rec}}$ are different and we can break the degeneracy between them in the parameter estimation although the degeneracy turns out to be rather strong.

\begin{figure}
\begin{center}
\begin{tabular}{cc}
\resizebox{80mm}{!}{\includegraphics{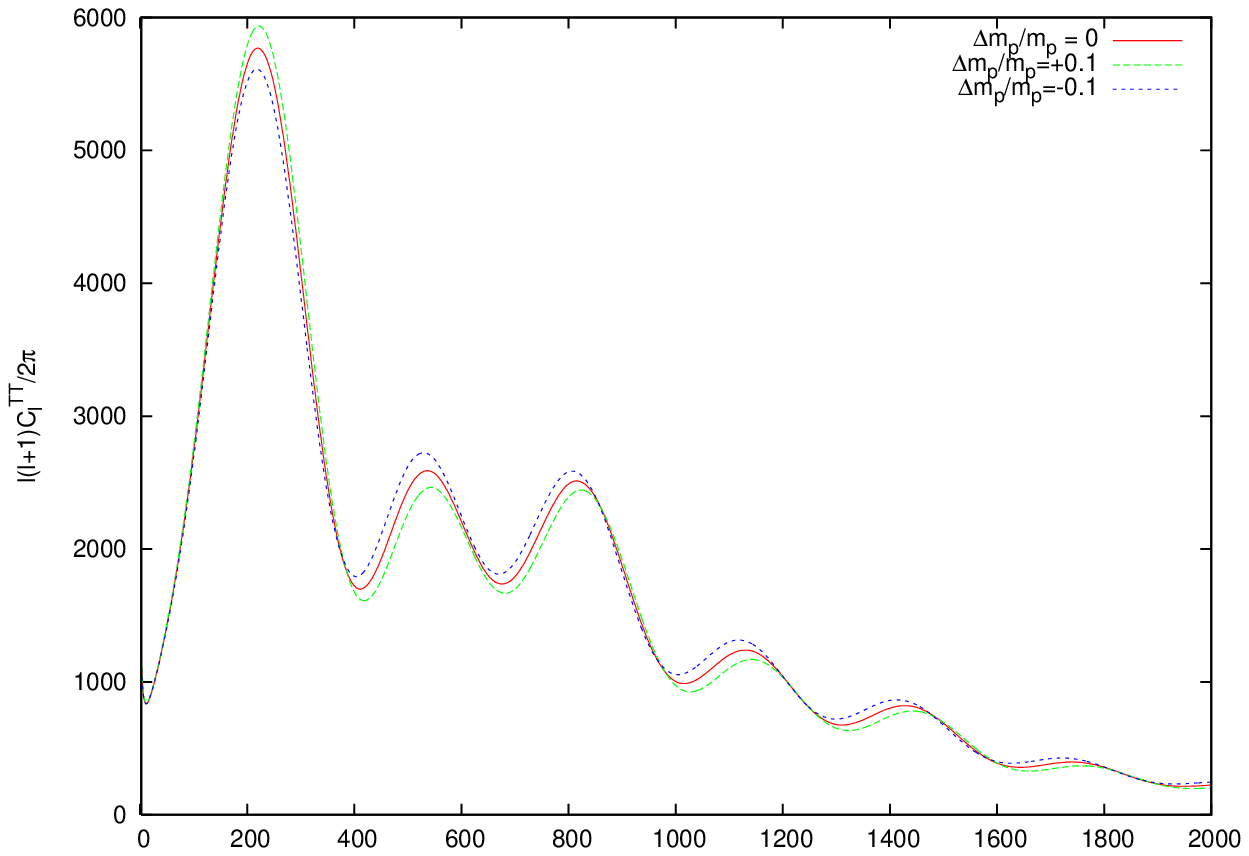}} &
\resizebox{80mm}{!}{\includegraphics{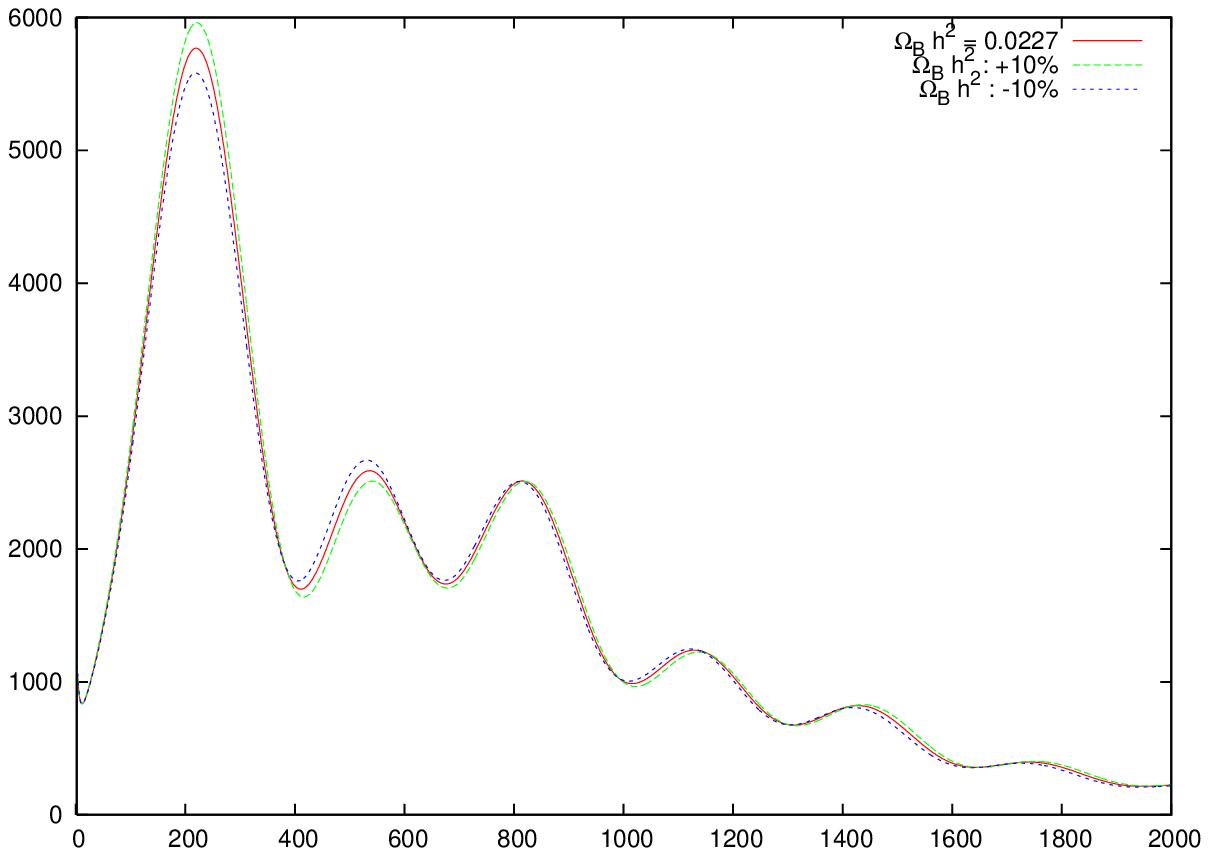}} \\
\resizebox{80mm}{!}{\includegraphics{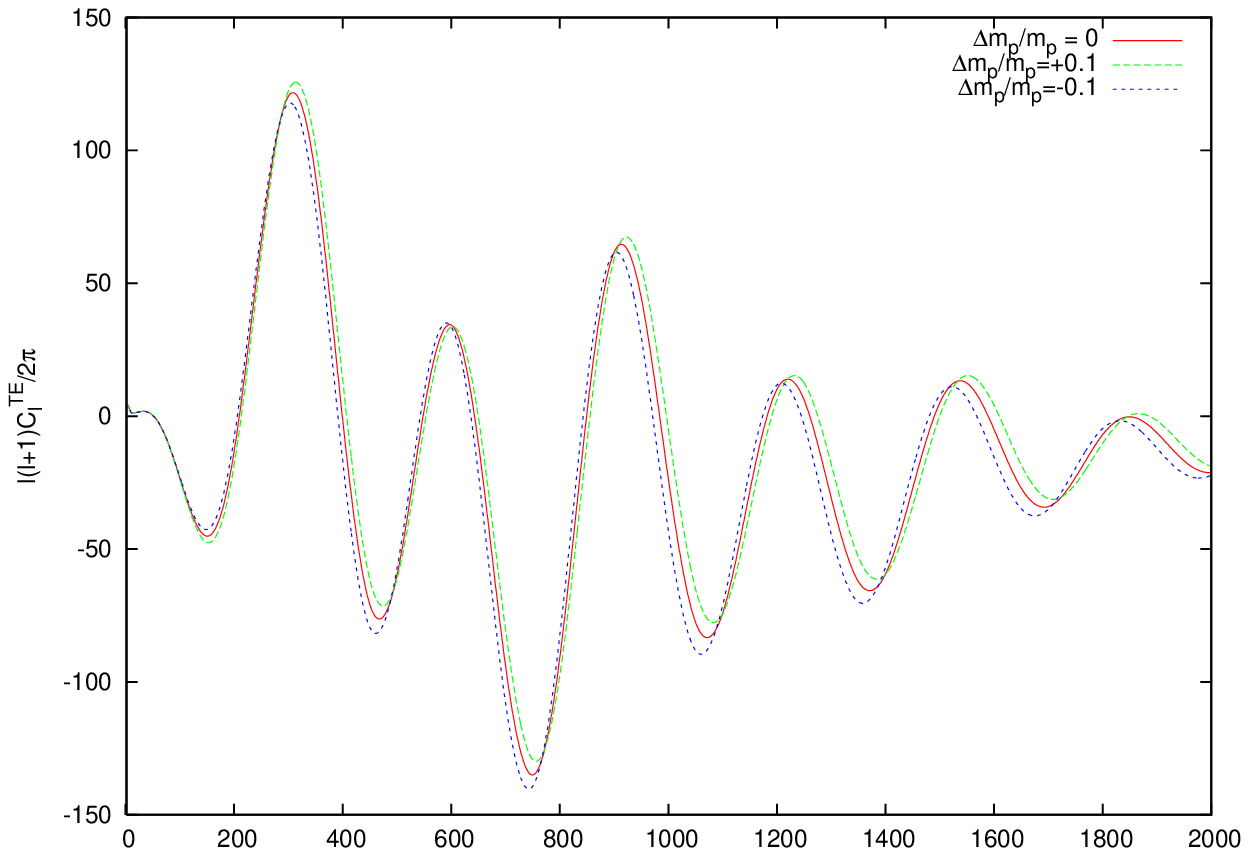}} &
\resizebox{80mm}{!}{\includegraphics{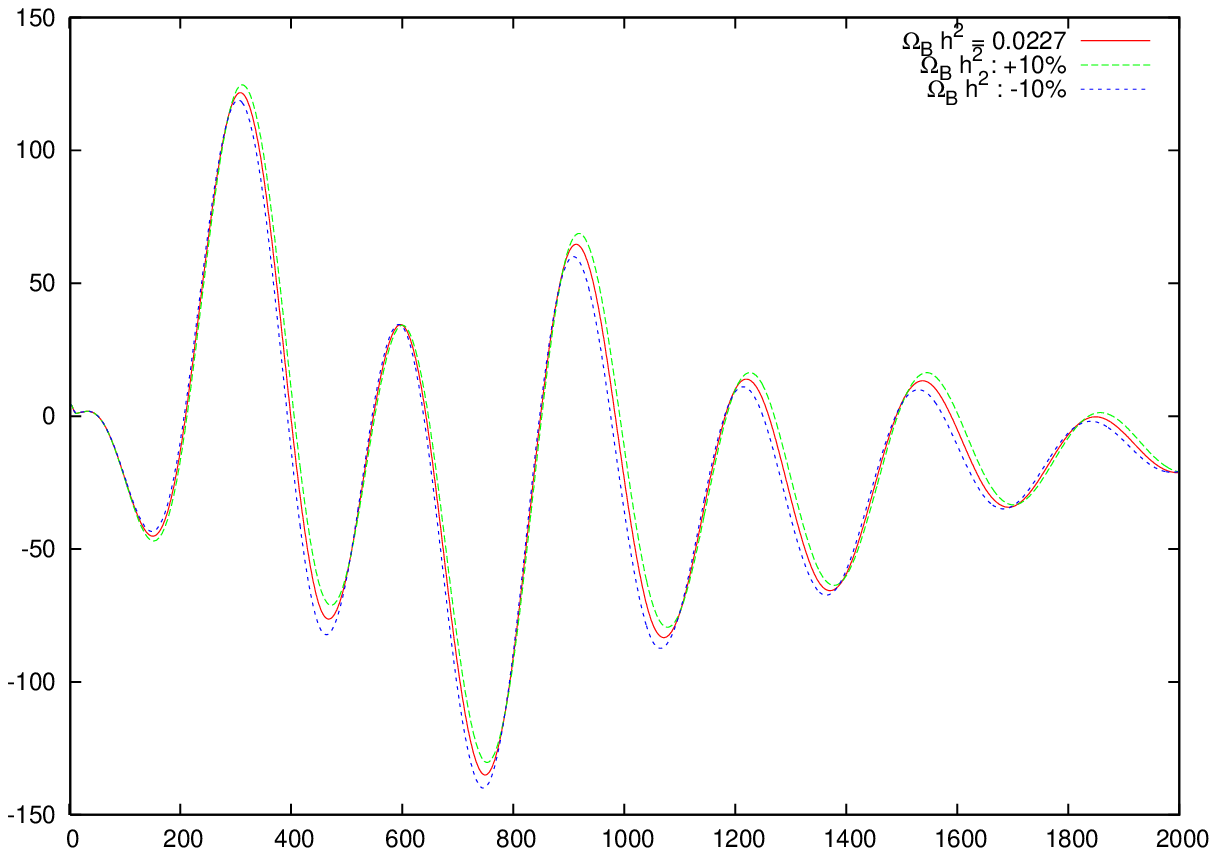}} \\
\resizebox{80mm}{!}{\includegraphics{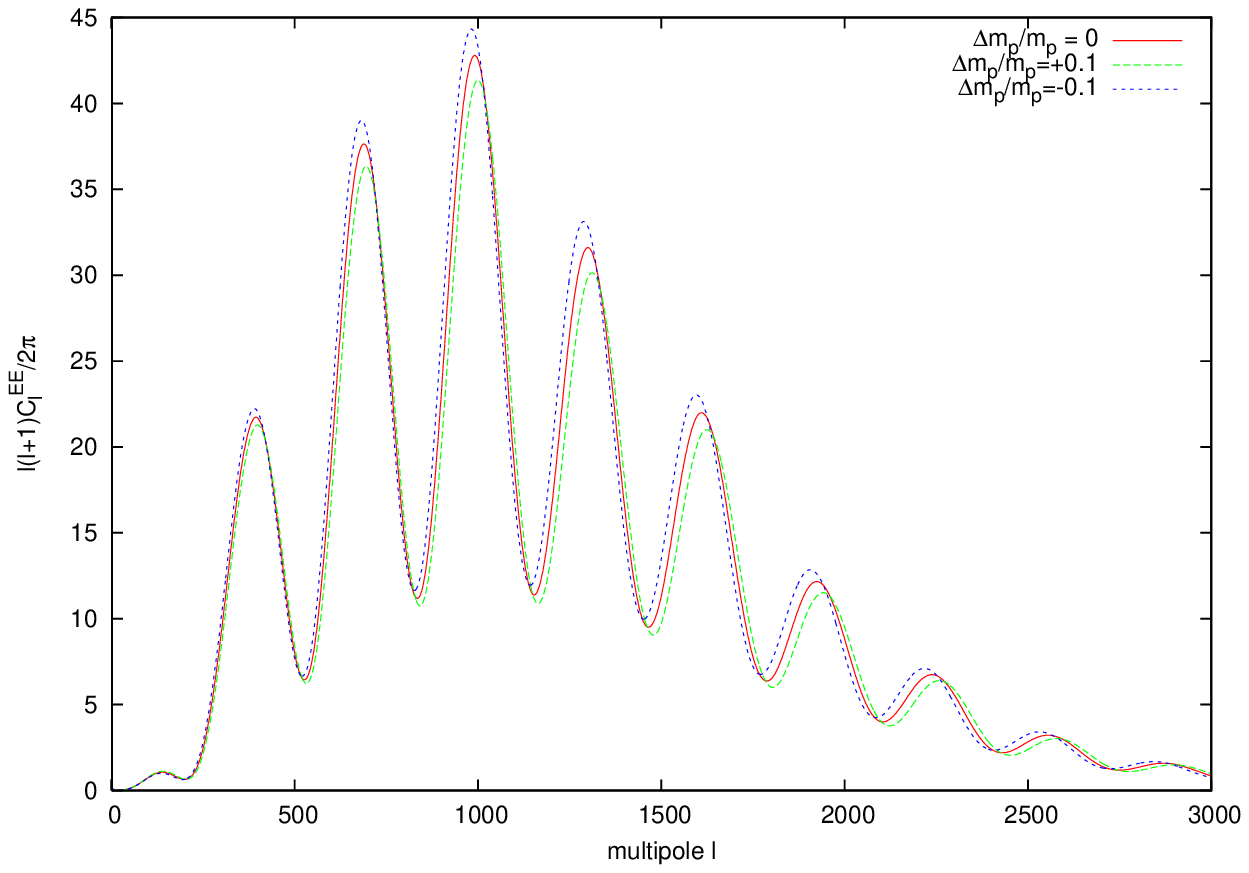}} &
\resizebox{80mm}{!}{\includegraphics{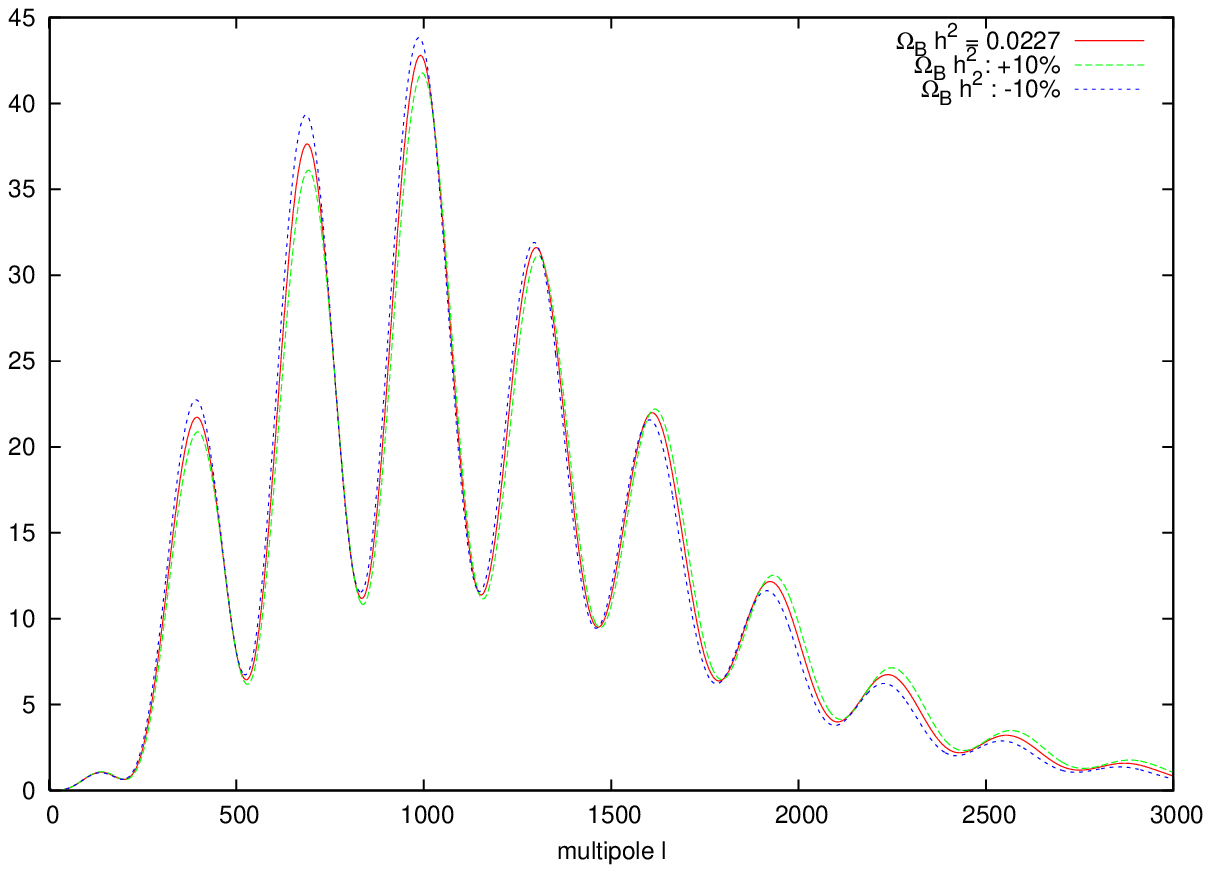}} \\
\end{tabular}
\caption{The top left panel is the CMB temperature anisotropy spectra for no change of $m_{p}$ (solid red curve), an increase of $m_{p}$ by 10$\%$ (dashed green curve), a decrease of $m_{p}$ by 10$\%$ (dotted blue curve). The middle left and bottom left panels are the TE and EE spectra respectively for three different values of $\Delta m_{p}/m_{p}$. For comparison, the corresponding spectra for three different values of $\Omega_{B}h^{2}$ are shown in the right hand side panels.}
\label{fig:wmap}
\end{center}
\end{figure}

With the careful attention to the difference between $\rho_{B}$ and $\Delta m_{p}$ noted in the last paragraph, we calculated the angular power spectra of the CMB temperature anisotropy, $C_{\ell}^{TT}$, for several different values of $\Delta m_{p}$ using the modified CAMB code \cite{camb}. The result is shown in FIG.~\ref{fig:wmap}, where we also draw the cross correlation of temperature and E-mode polarization, $C_{\ell}^{TE}$, and the angular power spectrum of E-mode polarization, $C_{\ell}^{EE}$, for completeness. Incorporating $\Delta m_{p}$ into the numerical calculation, we assume $m_{p}$ to change instantaneously at a specific time ($z=100$). In other words, $m_{p,\mathrm{rec}}$ was constant before $z=100$ and after that the proton mass becomes the present value 
\footnote{Here, only the mass difference $\Delta m_{p}$ matters and it is not important how the proton mass has changed between the recombination epoch and the present epoch, because characteristic signatures such as acoustic oscillation or Silk damping are imprinted before and during the recombination process, which is also true on the effects that $\Delta m_{p}$ induces.}. 
This procedure is adopted throughout this paper. We consider a flat $\Lambda$CDM universe with a power-law adiabatic primordial fluctuation and take the fiducial standard cosmological parameter values as the WMAP 5-year marginalized mean values \cite{Dunkley:2008ie,Komatsu:2008hk};
\begin{equation}
(\Omega_{B}h^{2}, \Omega_{DM}h^{2}, h, \tau, n_{s})=(0.0227, 0.1099, 0.72, 0.087, 0.963), \label{fiducial}
\end{equation}
where $\Omega_{B}$ is the baryon energy density $\rho_{B,\mathrm{0}}$ normalized by the critical density $\rho_{\mathrm{crit,0}}$, $\Omega_{DM}$ is the dark matter energy density $\rho_{DM,\mathrm{0}}$ normalized by the critical density $\rho_{\mathrm{crit,0}}$, $h$ is the current Hubble parameter, $H_{0}$, in units of 100\,km\,sec$^{-1}$\,Mpc$^{-1}$, $\tau$ is the optical depth to reionization and $n_{s}$ is the spectral index of the primordial curvature perturbation. As expected from the above argument, the effect of $\Delta m_{p}$ is very similar to that of $\Omega_{B}$. Due to the larger value of $\Delta m_{p}$, the height of the odd peaks increases compared with that of the even peaks, which is a very well-known result for $\Omega_{B}$ \cite{Hu:1995en}.

 Now we discuss how the variations of the parameters change the shape of the CMB power spectrum quantitatively. We use four quantities which have been proposed to characterize the temperature spectrum well \cite{Hu:2000ti}: the position of the first peak $\ell_{1}$, the height of the first peak relative to the large angular-scale amplitude evaluated at $\ell =10$,
\begin{equation}
H_{1}\equiv \left(\frac{\Delta T_{\ell_{1}}}{\Delta T_{10}}\right)^{2},
\end{equation}
the ratio of the second peak $\ell_{2}$ height to the first,
\begin{equation}
H_{2}\equiv \left(\frac{\Delta T_{\ell_{2}}}{\Delta T_{\ell_{1}}}\right)^{2},
\end{equation}
the ratio of the third peak $\ell_{3}$ height to the first,
\begin{equation}
H_{3}\equiv \left(\frac{\Delta T_{\ell_{3}}}{\Delta T_{\ell_{1}}}\right)^{2},
\end{equation}
where $(\Delta T_{\ell})^{2} \equiv \ell(\ell+1)C_{\ell}^{\mathrm{TT}}/2\pi$. We calculate the response of these four quantities when we vary the cosmological parameters. When we vary one parameter, the other parameters are fixed. The parameter set we consider is the standard flat-$\Lambda$CDM cosmological parameters plus varying physical constants: $\Omega_{B}h^{2}, \Omega_{m}h^{2}, h, \tau, n_{s}, \alpha_\mathrm{rec}, m_{e,\mathrm{rec}}, m_{p,\mathrm{rec}}$, where $\Omega_{m}$ means $\Omega_{B}+\Omega_{DM}$. The fiducial values of the standard cosmological parameters are given in (\ref{fiducial}), and those of physical constants equal to the present values. Let us define $\omega_{b}=\Omega_{B}h^{2},\omega_{m}=\Omega_{m}h^{2}$ and omit the subscript ``rec" of physical constants for simplicity, then we have found after extensive numerical calculations that the four characteristic measures of the angular power spectrum have the following dependence on the cosmological parameters and the physical constants.
\begin{align}
\Delta \ell_{1} &= 15.1 \frac{\Delta \omega_{b}}{\omega_{b}} -24.0\frac{\Delta \omega_{m}}{\omega_{m}} - 45.3 \frac{\Delta h}{h}+35.6 \frac{\Delta n_{s}}{n_{s}}-317 \frac{\Delta \alpha}{\alpha} +142 \frac{\Delta m_{e}}{m_{e}} + 19.1 \frac{\Delta m_{p}}{m_{p}}, \label{4v-1} \\ 
\Delta H_{1} &= 2.84 \frac{\Delta \omega_{b}}{\omega_{b}} -2.80 \frac{\Delta \omega_{m}}{\omega_{m}}  -2.02 \frac{\Delta h}{h}+18.3 \frac{\Delta n_{s}}{n_{s}}-0.64 \frac{\Delta \tau}{\tau} \notag \\
&  \ \ \ \ \ \ \ \ \ \ \ \ \ \ \ \ \ \ \ \ \ \ \ \ \ \ \ \ \ \ \ \ \ \ \ \ \ \ +4.42 \frac{\Delta \alpha}{\alpha} + 1.05 \frac{\Delta m_{e}}{m_{e}}+0.868 \frac{\Delta m_{p}}{m_{p}}, \label{4v-2} \\
\Delta H_{2} &= -0.29 \frac{\Delta \omega_{b}}{\omega_{b}} - 0.020 \frac{\Delta \omega_{m}}{\omega_{m}} +0.53  \frac{\Delta n_{s}}{n_{s}} + 0.75 \frac{\Delta \alpha}{\alpha} + 0.20  \frac{\Delta m_{e}}{m_{e}} -0.351 \frac{\Delta m_{p}}{m_{p}}, \label{4v-3} \\
\Delta H_{3} &= -0.18 \frac{\Delta \omega_{b}}{\omega_{b}} + 0.19 \frac{\Delta \omega_{m}}{\omega_{m}} +0.33  \frac{\Delta n_{s}}{n_{s}} + 0.42 \frac{\Delta \alpha}{\alpha} -0.008  \frac{\Delta m_{e}}{m_{e}} -0.246 \frac{\Delta m_{p}}{m_{p}}, \label{4v-4}
\end{align}
where the values at the fiducial parameter values are $\ell_{1}=219.9, H_{1}=6.864, H_{2} = 0.4482$ and $H_{3} =0.4342$, respectively. These formulae are extension of \cite{Ichikawa:2006nm} to incorporate the change of $m_{p}$. The coefficients of the $\Delta\omega_{b}/\omega_{b}$ and $\Delta m_{p}/m_{p}$ are very similar, which agrees with our qualitative discussion. Although two parameters $\omega_{b}$ and $ m_{p}$ have a strong degeneracy, it is not exact and we can expect that this small but finite difference actually enables us to discriminate these two parameters and give some limit on $\Delta m_{p}/m_{p}$. More quantitative analysis of the degeneracy is performed in the appendix.

\section{The Dilaton Dependence of $\alpha$, $m_{e}$ and $m_{p}$} \label{sec:dilaton}
In this section, we describe our setups for the dilaton dependence of the three physical constants and relation among them. This generally follows what is adopted in Refs.~\cite{Campbell:1994bf,Ichikawa:2002bt}. 
 As a concrete example, we consider the tree level low energy action of the heterotic string in the Einstein frame \cite{Campbell:1994bf,Lovelace:1983yv,Fradkin:1984pq,Callan:1985ia,Sen:1985eb,Sen:1985qt,Callan:1986jb}. The action is 
\begin{align}
S &= \int d^{4}x \sqrt{-g} \bigg(\frac{1}{2\kappa^{2}}R -\frac{1}{2}\partial_{\mu}\Phi\partial^{\mu}\Phi-\frac{1}{2}D_{\mu}\phi D^{\mu}\phi \notag \\
 & \ \ \ \ \ \ -\Omega^{-2}V(\phi) -\bar{\psi}\gamma_{\mu}D^{\mu}\psi- \Omega^{-1}m_{\psi}\bar{\psi}\psi-\frac{\alpha'}{16\kappa^{2}}\Omega^{2} F_{\mu\nu}F^{\mu\nu}\bigg), \label{sayou}
\end{align}
where $\kappa^{2}=8\pi G=8\pi/M_{pl}^{2}$, where $M_{pl}$ is the Planck scale, $\Phi$ is the dilaton field, $\phi$ is an arbitary scalar field, $\psi$ is an arbitary fermion, $D_{\mu}$ is the gauge covariant derivative corresponding to gauge fields with field strength $F_{\mu\nu}$, and $F_{\mu\nu}$ is the gauge field with gauge group including $\mathrm{SU(3)}\times\mathrm{SU(2)}\times\mathrm{U(1)}$. $\Omega$ is the conformal factor used to transfer from the string frame and defined as 
\begin{equation}
\Omega = e^{-\kappa \Phi/\sqrt{2}}.
\end{equation}

Comparing the gauge field strength term $-(\alpha'/16\kappa^{2})\Omega^{2}F_{\mu\nu}F^{\mu\nu}$ of the action (\ref{sayou}) with the definition of the Lagrangian density $-(1/4g^{2})F_{\mu\nu}F^{\mu\nu}$ where $g$ is the unified coupling constant, we get 
\begin{equation}
\frac{1}{g^{2}(M_{pl})}=\frac{\alpha^{'} e^{-\sqrt{2}\kappa \Phi}}{4\kappa^{2}}\equiv \frac{\alpha^{'}S}{4\kappa}.
\end{equation}
The gauge coupling constants at low energy scale are calculated using renormalization group equations. As for the fine structure constant $\alpha$, we can check that $\alpha$ does not run practically, so $\alpha$ at low energy (or CMB energy scale which is equal to the energy scale at which recombination took place) is determined in good approximation as 
\begin{equation}
\alpha_{\mathrm{rec}} \approx \alpha(M_{pl})=\frac{g(M_{pl})^{2}}{4\pi}=\frac{\kappa^{2}}{\pi \alpha^{'}}e^{\sqrt{2}\kappa\Phi}.
\end{equation} 
From this expression, our variable $\Delta\alpha/\alpha$ becomes
\begin{equation}
\frac{\Delta\alpha}{\alpha} = \frac{\alpha_{\mathrm{rec}}-\alpha_{\mathrm{0}}}{\alpha_{\mathrm{0}}}=e^{\sqrt{2}\kappa\Delta\Phi}-1 \label{alpha_dil}
\end{equation}
where $\Delta\Phi = \Phi_{\mathrm{rec}}-\Phi_{\mathrm{0}}$. 

From the solution of the one-loop renormalization group equation for the $\mathrm{SU(3)}$ coupling constant $g_{3}$\footnote{The integration constant is determined by $g_{3}(\Lambda_{\mathrm{QCD}})=\infty$}, 
\begin{equation}
\frac{\kappa^{2}}{\pi \alpha^{'}}e^{\sqrt{2}\kappa\Phi} = \frac{g_{3}(M_{pl})^{2}}{4\pi} \simeq \frac{12\pi}{27 \log (M_{pl}^{2}/\Lambda_{\mathrm{QCD}}^{2})}. 
\end{equation}
Using this relation and the fact that $g(M_{pl})^{2}=0.1$ and that the proton mass $m_{p}$ is proportional to the QCD energy scale $\Lambda_{\mathrm{QCD}}$, $\Delta m_{p}/m_{p}$ is also expressed through $\Delta\alpha/\alpha$ as
\begin{equation}
\frac{\Delta m_{p}}{m_{p}} \equiv \frac{m_{p,\mathrm{rec}}-m_{p,\mathrm{now}}}{m_{p,\mathrm{now}}} = \frac{\Delta \Lambda_{\mathrm{QCD}}}{\Lambda_{\mathrm{QCD}}} =\exp \left( \frac{80\pi^{2}}{9}\frac{\frac{\Delta\alpha}{\alpha}}{1+\frac{\Delta\alpha}{\alpha}} \right) -1. \label{mp_change}
\end{equation}
This relation is peculiar in that a small increase in $\Delta\alpha/\alpha$ results in a large increase in $\Delta m_{p}/m_{p}$, which has a great impact on the parameter estimation.

Finally, we can read the electron mass term from the action as $e^{\kappa\Phi\sqrt{2}}m_{\psi}\bar{\psi}\psi$. Assuming that the Higgs vacuum expectation value $\langle H\rangle=\mathrm{const.}$, the time variation of $m_{e}$ is expressed as   
\begin{equation}
\frac{\Delta m_{e}}{m_{e}}=\frac{m_{e,\mathrm{rec}}-m_{e,\mathrm{0}}}{m_{e,\mathrm{0}}} = e^{\kappa\Delta\Phi/\sqrt{2}}=\sqrt{1+\frac{\Delta\alpha}{\alpha}}-1. \label{me_change}
\end{equation}

\section{Constraints on the variation of the coupling constants} \label{sec:constraint}

\begin{figure}
\centering
\includegraphics[width=16cm]{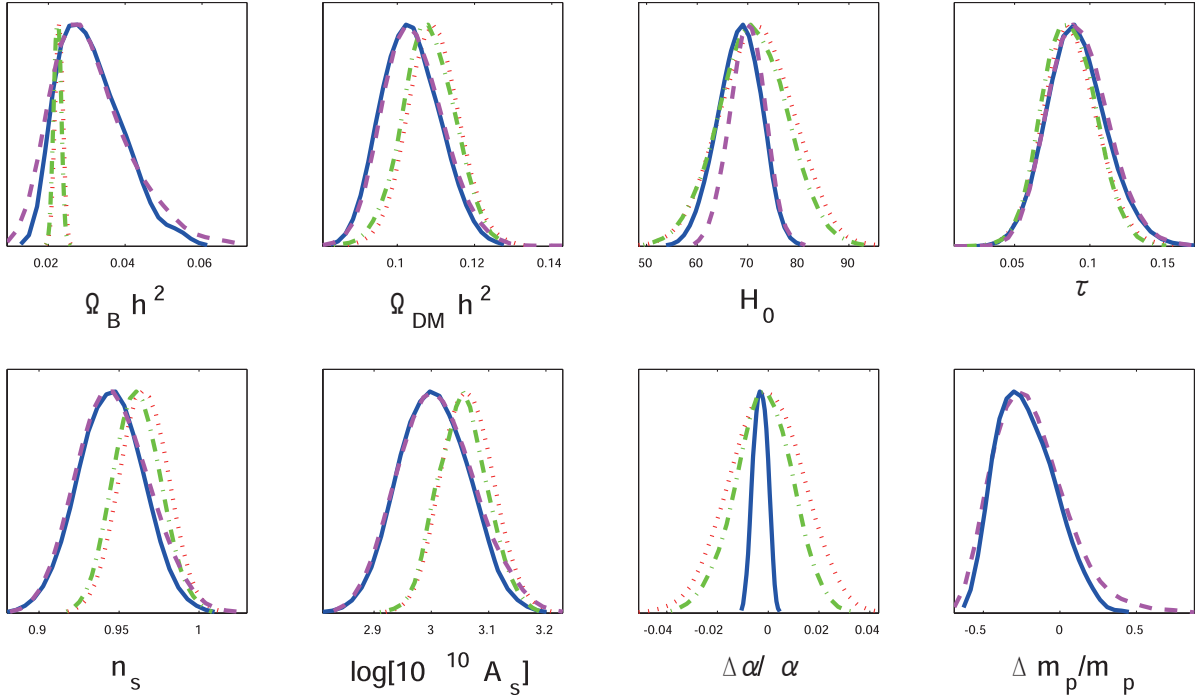}
\caption{One-dimensional marginalized posterior distributions for four models: varying all three constants $\alpha, m_{e}$ and $m_{p}$ assuming the relation (\ref{mp_change}) and (\ref{me_change}) (solid blue curves), varying only two constants $\alpha$ and $m_{e}$ assuming the relation (\ref{me_change}) (dot-dashed green curves), varying only one constant $\alpha$ (dotted red curves) and varying only one constant $m_{p}$ (dashed magenta curves).}
\label{fig:one-d-three}
\end{figure}

\begin{table}
\caption{MCMC results on the mean values and 68$\%$ confidence intervals of cosmological parameters for four versions: varying $\alpha$ only, varying $\alpha$ and $m_{e}$, varying $\alpha, m_{e}$ and $m_{p}$, varying $m_{p}$ only.}
\label{hyou1}
\begin{center}
\begin{tabular}{|l|c|c @{\vrule width 1.3pt\ }c @{\ \vrule width 1.3pt} c|}
\hline 
  & varying $\alpha$ only & varying $\alpha$ and $m_{e}$ & varying $\alpha, m_{e}$ and $m_{p}$ & varying $m_{p}$ only \\
\hline \hline
$100\Omega_{B}h^{2}$ & $2.27^{+0.10}_{-0.10}$ & $2.24^{+0.09}_{-0.09}$ & $3.13^{+0.85}_{-0.83}$ & $3.13^{+0.97}_{-0.95}$ \\ \hline
$\Omega_{DM}h^{2}$ & $0.109^{+0.006}_{-0.006}$ & $0.108^{+0.007}_{-0.006}$ & $0.103^{+0.008}_{-0.007}$ & $0.104^{+0.007}_{-0.007}$ \\ \hline
$\tau$ & $0.0877^{+0.007}_{-0.008}$ & $0.0864^{+0.008}_{-0.009}$ & $ 0.0916^{+0.007}_{-0.010}$ & $0.0928^{+0.008}_{-0.010}$   \\ \hline
$n_{s}$ & $0.966^{+0.014}_{-0.015}$ & $0.961^{+0.014}_{-0.015}$ & $0.945^{+0.020}_{-0.020}$ & $0.947^{+0.022}_{-0.022}$  \\ \hline 
$\log(10^{10} A_{s})$ & $3.06^{+0.05}_{-0.04}$ & $3.06^{+0.04}_{-0.04}$ & $3.00^{+0.06}_{-0.06}$ & $3.01^{+0.06}_{-0.07}$ \\ \hline
$H_{0}$ & $71.9^{+7.0}_{-7.1}$ & $71.0^{+6.4}_{-6.4}$ & $68.2^{+4.2}_{-4.2}$ & $69.9^{+3.3}_{-3.4}$ \\ \noalign{\hrule height 1.3pt}
$\Delta\alpha/\alpha$ & $-0.00089^{+0.0131}_{-0.0148}$ & $-0.0019^{+0.011}_{-0.011}$ & $-0.0031^{+0.0028}_{-0.0029}$ & -- \\ 
\noalign{\hrule height 1.3pt}
$\Delta m_{p}/m_{p}$ & -- & -- & $-0.22^{+0.19}_{-0.19}$ & $-0.19^{+0.22}_{-0.23}$ \\ 
\hline
\end{tabular}
\end{center}
\end{table}

Based on the model described in the previous section, we constrain the variation of $\alpha$ using three kinds of CMB anisotropy spectra, $C_{\ell}^{TT}$, $C_{\ell}^{EE}$, and $C_{\ell}^{TE}$ of the five-year WMAP data \cite{Nolta:2008ih,Dunkley:2008ie,Komatsu:2008hk}. Theoretical anisotropy spectra are calculated by the CAMB code \cite{Lewis:1999bs,camb} modified to include the varying physical constants as in Sec.~\ref{sec:effect}. We performed the parameter estimation using Markov-Chain Monte-Carlo (MCMC) techniques implemented in the public CosmoMC code \cite{Lewis:2002ah,cosmomc}. 

 We have run the CosmoMC code on eight Markov chains in each case. To check the convergence, we used the ``variance of chain means"/``mean of chain variances" $R$ statistic and adopted the condition $R-1 < 0.03$. In all the analysis below, we have also incorporated the result of Hubble Key Project of the Hubble Space Telescope (HST) on the Hubble parameter $H_{0}$, which means that we have imposed Gaussian prior of $H_{0} = 72\pm 8\,\mathrm{km\ sec^{-1} \ Mpc^{-1}}$ \cite{Freedman:2000cf}. In our analysis, we derive the confidence limits for standard parameters $(\Omega_{B}h^{2},\Omega_{DM} h^{2},H_{0},n_{s},A_{s},\tau)$ plus some varying physical constants.  

 Figures~\ref{fig:one-d-three} show 1D-marginalized posterior distributions of the eight parameters for the case not only $\alpha$ but also $m_{e}$ and $m_{p}$ are changed simultaneously according to the model described in the previous section. This is the main result of this paper and for comparison with the literature we have also plotted the distributions for two other cases: the case of varying only $\alpha$ \cite{Nakashima:2008cb} and the case of varying both $\alpha$ and $m_{e}$ assuming the relation (\ref{me_change}) which has already been analyzed in \cite{Scoccola:2008jw}. In TABLE \ref{hyou1}, we also present the mean values and the 68$\%$ confidence intervals of the cosmological parameters for these three cases. We can conclude from these results that the inclusion of the variation of $m_{p}$ results in a tighter limit on the variation of $\alpha$, which is now  
\begin{equation}
-8.28\times 10^{-3} < \frac{\Delta\alpha}{\alpha} < 1.81 \times 10^{-3}, \label{mainresult}
\end{equation}
compared with the limit for the two other cases: $-0.024 < \Delta\alpha/\alpha < 0.019$ (for varying $\alpha, m_{e}$) and $-0.028 < \Delta\alpha/\alpha < 0.026$ (for vayring only $\alpha$), all of which are at 95$\%$C.L.

\begin{figure}
\centering
\includegraphics[width=12cm]{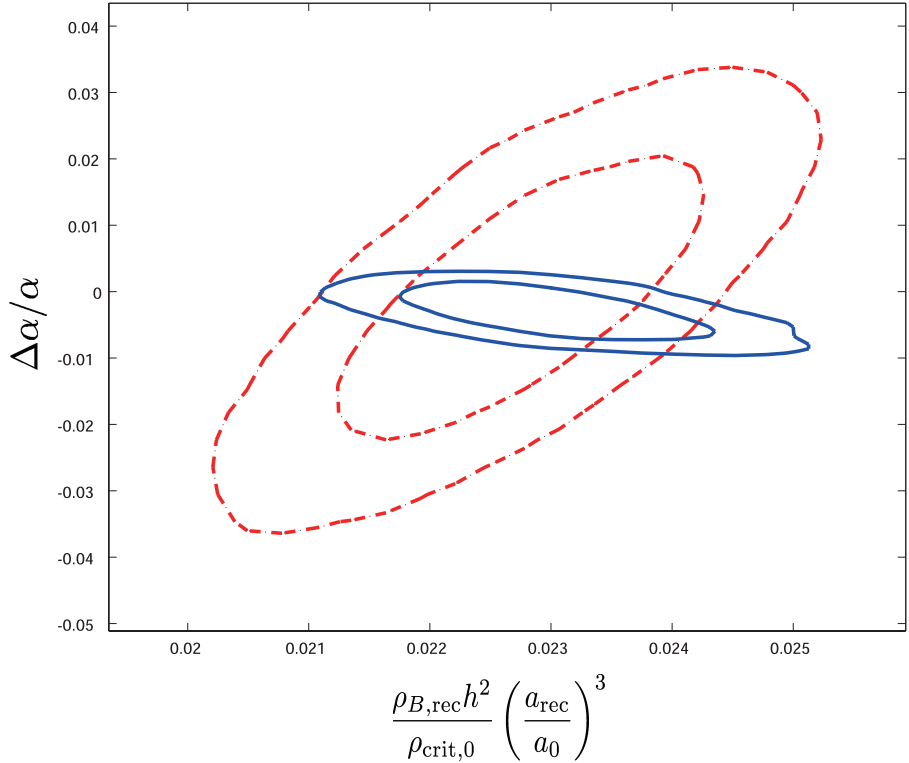}
\caption{Two-dimensional marginalized posterior distributions of $\rho_{B,\mathrm{rec}}h^{2}/\rho_{\mathrm{crit,0}}$ and $\Delta\alpha/\alpha$ for two models: varying all three constants $\alpha, m_{e}$ and $m_{p}$ assuming the relation (\ref{mp_change}) and (\ref{me_change}) (solid blue contours) and varying only one constant $\alpha$ (dot-dashed red contours). The plotted $\rho_{B,\mathrm{rec}}h^{2}/\rho_{\mathrm{crit,0}}$ (horizontal axis) is actually $(\rho_{B,\mathrm{rec}}h^{2}/\rho_{\mathrm{crit,0}})( a_{\mathrm{rec}}/a_{\mathrm{0}})^{3}$, which corresponds to $(\rho_{B,\mathrm{0}}h^{2}/\rho_{\mathrm{crit,0}})(1+\Delta m_{p}/m_{p})$ for the former case and $\rho_{B,\mathrm{0}}h^{2}/\rho_{\mathrm{crit,0}}$ for the latter case. In each pair of contours, the inner one shows 68$\%$C.L. and the outer one shows 95$\%$C.L.}
\label{fig:ombz}
\end{figure}

 In FIG.~\ref{fig:one-d-three} and TABLE \ref{hyou1}, we can also see that, as a cost of tightening of the constraint on $\alpha$, the limit on $\Omega_{B}h^{2}$, which is one of the most tightly constrained parameter from CMB data in ordinary analysis, become very loose because of the degeneracy between $\Omega_{B}$ and $\Delta m_{p}/m_{p}$. This result is reasonable from the argument in Sec.~\ref{sec:effect}. On the other hand, since CMB is sensitive to the values of cosmological parameters at the recombination epoch, the limit on $\Omega_{B,\mathrm{rec}}$ in the present scenario is expected to be of the same order as that on $\Omega_{B,\mathrm{rec}}$ in the case without $m_{p}$ variation. We can confirm  it from FIG.~\ref{fig:ombz}, which shows 2D-marginalized posterior distributions of $\Delta\alpha/\alpha$--$\rho_{B,\mathrm{rec}}h^{2}/\rho_{\mathrm{crit,0}}$ for two cases, one with both $m_{p}$ and $m_{e}$ variation and the other with no variation in $m_{p}$ and $m_{e}$.  

 One of the reasons why we could obtain tighter constraint than the case without varying $m_{p}$ is thought to be the specific relation (\ref{mp_change}) in our model. To ensure that statement, we have also implemented MCMC calculations in the case that only $m_{p}$ is varied and the other two physical constants remain constant, which means that we do not impose any model in this case. The result is also shown in FIG.~\ref{fig:one-d-three} as 1D-marginalized posterior distributions. Two marginalized curves in the panel showing $\Delta m_{p}/m_{p}$ are very similar and those limits at 95$\%$C.L. are $-0.54 < \Delta m_{p}/m_{p} < 0.36$ (for varying $m_{p}$ only) and $-0.51 < \Delta m_{p}/m_{p} < 0.17$ (for varying $\alpha, m_{e}$ and $m_{p}$), which implies that the dominant factor in constraining the variation of physical constants under the relation (\ref{me_change}) and (\ref{mp_change}) is $\Delta m_{p}/m_{p}$ and it is the specific relation between $\Delta\alpha/\alpha$ and $\Delta m_{p}/m_{p}$ that determines the order of magnitude of the resultant $\Delta\alpha/\alpha$ limit in the case these parameters are varied simultaneously through the dilaton's motion. 

\begin{figure}
\centering
\includegraphics[width=12cm]{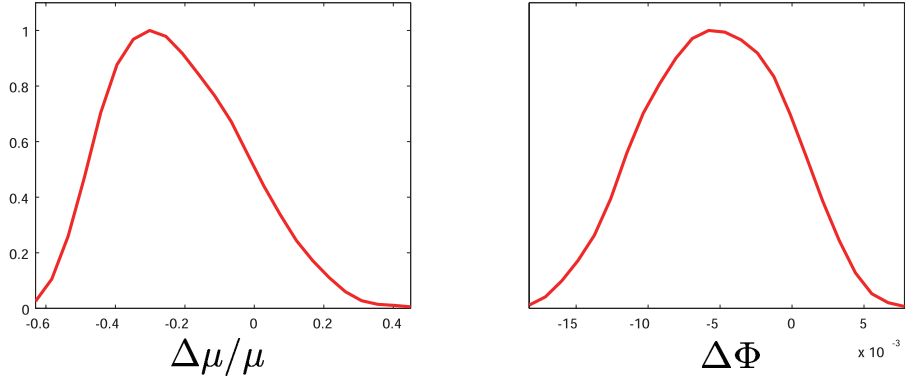}
\caption{One-dimensional marginalized posterior distributions for two parameters: $\Delta \mu/\mu$ and $\Delta\Phi$}
\label{fig:mu_phi}
\end{figure}

 Finally, in our model, we can calculate the limit on not only $\Delta\alpha/\alpha$ but also $\Delta\Phi$ and $\Delta\mu/\mu$ where $\mu\equiv m_{p}/m_{e}$. The results are shown in FIG.~\ref{fig:mu_phi}. $\Delta\Phi$ is the dilaton field variation and using the relation (\ref{alpha_dil}) with $\kappa =\sqrt{8\pi G}=(2.43\times10^{18} [\mathrm{GeV}])^{-1}$, we obtain at 95$\%$C.L.
\begin{equation}
-1.4\times 10^{16} [\mathrm{GeV}] < \Delta\Phi <3.1 \times 10^{15} [\mathrm{GeV}].
\end{equation}
On the other hand, the constraint on $\Delta\mu/\mu$ is, at 95$\%$C.L., 
\begin{equation}
-0.52 < \frac{\Delta\mu}{\mu} < 0.17. 
\end{equation}
Note that, although it is not so stringent compared with those obtained from other experiments \cite{Reinhold:2006zn}, constraining this parameter in the recombination epoch is meaningful because in the previous works the variation of $m_{p}$ has never been considered. We also expect that future CMB measurement such as PLANCK will give more stringent limits of these parameters (see Appendix B).

\section{Conclusion} \label{sec:conclusion}

In summary, we have pointed out that the time variation of $\Lambda_{\mathrm{QCD}}$ can affect the CMB through the variation in the proton mass. We have shown that there is a notable effect when we impose the widely adopted assumptions of unification models found in the literature \cite{Campbell:1994bf}. From the WMAP 5-year data, by MCMC analysis using the CosmoMC code, we have obtained the new and stringent constraint on the time variation of the fine structure constant as $-8.28\times 10^{-3}<\Delta\alpha/\alpha<1.81\times 10^{-3}$. We have also verified that, in getting the above result, the newly considered effect of $\Delta m_{p}/m_{p}$ is very important in that the degeneracy between $\Delta m_{p}/m_{p}$ and $\Omega_{B}$ is tremendously strong and that the small variation of $\Delta\alpha/\alpha$ leads to the large variation of $\Delta m_{p}/m_{p}$. Furthermore, by including the varying $m_{p}$, we have obtained a limit on the variation of proton-electron mass ratio $\mu$ as $-0.52 < \Delta\mu/\mu < 0.17$. 

CMB observation will be much more precise to give a lot of information around the damping tail of the power spectrum in near future. The difference in the CMB power spectra between $\Delta m_{p}/m_{p}$ and $\Omega_{B}$ appears in the small scale diffusion damping because the damping scale is determined by the baryon number density $n_{B}$, probing the time variation of coupling constants including $\Lambda_{\mathrm{QCD}}$ by CMB will be even more promising. This is discussed more quantitatively in appendix A, and supported by a forecast for future CMB survey in appendix B.

\bigskip

\section*{Acknowledgements}
We would like to thank Asantha Cooray for useful comments and Toyokazu Sekiguchi for useful help. M.N. is supported by JSPS through research fellowships. This work was partially supported by JSPS Grant-in-Aid for Scientific Research No.~19340054 and Global COE Program "The Physical Sciences Frontier," MEXT, Japan.

\section*{Appendix A}
In this appendix, we discuss the degeneracies between cosmological parameters using the linear algebra. In section \ref{sec:effect}, we have introduced four variables $\ell_{1}, H_{1}, H_{2}, H_{3}$ and calculated how they depend on the cosmological parameters and the physical constants $(\omega_{b}, \omega_{m}, h, \tau, n_{s}, \alpha_{\mathrm{rec}}, m_{e,\mathrm{rec}}, m_{p,\mathrm{rec}})$, the results of which are shown in (\ref{4v-1})-(\ref{4v-4}). These are rewritten in the matrix form as follows. 
\begin{equation}
\left(
\begin{array}{c}
\frac{\Delta \ell_{1}}{\ell_{1}} \\
\frac{\Delta H_{1}}{H_{1}} \\
\frac{\Delta H_{2}}{H_{2}} \\
\frac{\Delta H_{3}}{H_{3}} 
\end{array}
\right)=\left(
\begin{array}{cccccccc}
0.069 & -0.109 & -0.206 & 0.162 & 0 & -1.442 & 0.646 & 0.087 \\
0.414 & -0.408 & -0.294 & 2.666 & -0.093 & 0.644 & 0.153 & 0.127 \\
-0.647 & -0.045 & 0 & 1.183 & 0 & 1.673 & 0.446 & -0.783 \\
-0.415 & 0.438 & 0 & 0.760 & 0 & 0.967 & -0.018 & -0.567 
\end{array}
\right) \left(
\begin{array}{c}
\frac{\Delta \omega_{b}}{\omega_{b}} \\
\frac{\Delta \omega_{m}}{\omega_{m}} \\
\frac{\Delta h}{h} \\
\frac{\Delta n_{s}}{n_{s}} \\
\frac{\Delta \tau}{\tau} \\
\frac{\Delta \alpha}{\alpha} \\
\frac{\Delta m_{e}}{m_{e}} \\
\frac{\Delta m_{p}}{m_{p}} \\
\end{array}
\right), \label{append_eq1}
\end{equation}
where we have divided the both sides of each equation by the fiducial values $\ell_{1} = 219.9, H_{1} = 6.864, H_{2}=0.4482, H_{3} = 0.4342$, respectively. The $4\times 8$ matrix connects the cosmological parameters and the CMB variables. Such a rectangular $n\times m$ matrix, which we call $A$, can be decomposed in general as 
\begin{equation}
A=U\Sigma V^{\mathrm{T}}\ \ \ (U:n\times n,\ \Sigma :n\times m,\ V:m\times m), 
\end{equation}
which is known as the singular value decomposition. If $n < m$ and $\mathrm{rank}A =n$, then the form of $\Sigma$ is like
\begin{equation}
\Sigma = \left( 
\begin{array}{ccc|c}
\sigma_{1} & & &  \\
 & \ddots & & 0\\
 & & \sigma_{n}&
\end{array}
\right),  
\end{equation}  
where $\sigma_{i}$ is the non-zero singular value of the matrix $A$ and the matrices $U$ and $V$ are orthonormal ones. Applying the singular value decomposition to the $4\times 8$ matrix in (\ref{append_eq1}) and using the notation $D=U\Sigma$, we obtain
\begin{align}
D&=\left(
\begin{array}{cccccccc}
-0.672 & -1.227 & -0.794 & 0.033 & 0 & 0 & 0 & 0 \\
2.463 & -1.350 & 0.322 & 0.008 & 0 & 0 & 0 & 0 \\
2.1145 & 0.815 & -0.488 & -0.238 & 0 & 0 & 0 & 0 \\
1.268 & 0.614 & -0.232 & 0.398 & 0 & 0 & 0 & 0 
\end{array}
\right) \\
V^{\mathrm{T}}&=\left(
\begin{array}{cccccccc}
-0.073 & -0.037 & -0.046 & 0.788 & -0.018 & 0.581 & 0.069 & -0.168 \\
-0.326 & 0.209 & 0.149 & -0.542 & 0.029 & 0.654	& -0.148 & -0.289 \\
0.478 & -0.121 & 0.067 & -0.024 & -0.029 & 0.304 & -0.660 & 0.473 \\
-0.028 & 0.825 & -0.042 & 0.222 & -0.003 & -0.253 & -0.419 & -0.166 \\
0.019 &	-0.013 & 0.027 & 0.0350 & 0.998 & -0.006 & -0.020 & 0.013 \\
0.702 &	0.331 &	-0.318 & -0.169 & 0.018 & 0.230 & 0.450 & -0.106 \\
0.0670 & 0.285 & 0.817 & 0.074 & -0.019 & 0.049 & 0.347 & 0.344 \\
-0.402 & 0.261 & -0.447 & -0.028 & 0.019 & 0.153 & 0.196 & 0.712
\end{array}
\right)
\end{align}
Now, (\ref{append_eq1}) becomes
\begin{equation}
\left(
\begin{array}{c}
\frac{\Delta \ell_{1}}{\ell_{1}} \\
\frac{\Delta H_{1}}{H_{1}} \\
\frac{\Delta H_{2}}{H_{2}} \\
\frac{\Delta H_{3}}{H_{3}} 
\end{array}
\right) = DV^{\mathrm{T}}\left(
\begin{array}{c}
\frac{\Delta \omega_{b}}{\omega_{b}} \\
\frac{\Delta \omega_{m}}{\omega_{m}} \\
\frac{\Delta h}{h} \\
\frac{\Delta n_{s}}{n_{s}} \\
\frac{\Delta \tau}{\tau} \\
\frac{\Delta m_{p}}{m_{p}} 
\end{array}
\right) \approx D\left(
\begin{array}{c}
0.788 \frac{\Delta n_{s}}{n_{s}}+0.581 \frac{\Delta\alpha}{\alpha} \\
-0.542 \frac{\Delta n_{s}}{n_{s}}+0.654 \frac{\Delta\alpha}{\alpha} \\
0.478 \frac{\Delta \omega_{b}}{\omega_{b}}-0.660 \frac{\Delta m_{e}}{m_{e}} +0.473 \frac{\Delta m_{p}}{m_{p}} \\
0.825 \frac{\Delta\omega_{m}}{\omega_{m}}-0.419 \frac{\Delta m_{e}}{m_{e}} \\
0.999 \frac{\Delta \tau}{\tau} \\
-0.702 \frac{\Delta \omega_{b}}{\omega_{b}}+0.451 \frac{\Delta m_{e}}{m_{e}} \\
0.817 \frac{\Delta h}{h} \\
-0.402 \frac{\Delta \omega_{b}}{\omega_{b}}-0.447 \frac{\Delta h}{h}+0.712 \frac{\Delta m_{p}}{m_{p}} 
\end{array}
\right)
\end{equation}
At the last transformation, we have neglected the terms whose coefficients are small and irrelevant. Remembering the explict form of the matrix $D$ that the last four columns are zero, the last four components in the column vector represent the four degeneracy directions, $0.999\Delta\tau/\tau, -0.702\Delta\omega_{b}/\omega_{b}+0.451\Delta m_{e}/m_{e}, 0.817\Delta h/h, -0.402\Delta \omega_{b}/\omega_{b}-0.447\Delta h/h+0.712\Delta m_{p}/m_{p}$. 

In the above case, where all the variations of physical constants are included in the analysis, the final results are complicated and it is difficult to understand the degeneracies cleary. Therefore, we consider simpler cases in order now.

First, we treat only standard cosmological parameters, so (\ref{append_eq1}) is reduced to 
\begin{equation}
\left(
\begin{array}{c}
\frac{\Delta \ell_{1}}{\ell_{1}} \\
\frac{\Delta H_{1}}{H_{1}} \\
\frac{\Delta H_{2}}{H_{2}} \\
\frac{\Delta H_{3}}{H_{3}} 
\end{array}
\right)=\left(
\begin{array}{cccccc}
0.069 & -0.109 & -0.206 & 0.162 & 0 \\
0.414 & -0.408 & -0.294 & 2.666 & -0.093 \\
-0.647 & -0.045 & 0 & 1.183 & 0 \\
-0.415 & 0.438 & 0 & 0.760 & 0  
\end{array}
\right) \left(
\begin{array}{c}
\frac{\Delta \omega_{b}}{\omega_{b}} \\
\frac{\Delta \omega_{m}}{\omega_{m}} \\
\frac{\Delta h}{h} \\
\frac{\Delta n_{s}}{n_{s}} \\
\frac{\Delta \tau}{\tau} 
\end{array}
\right).
\end{equation}
The singular value decomposition of the $4\times 5$ matrix gives 
\begin{equation}
D=\left(
\begin{array}{ccccc}
-0.190 & 0.131 & -0.052 & -0.172 & 0 \\
-2.713 & 0.420 & 0.039 & 0.017 & 0 \\
-1.171 & -0.626	& -0.239 & 0.006 & 0 \\
-0.708 & -0.609 & 0.261 & -0.031 & 0
\end{array}
\right), V^{\mathrm{T}}=
\left(
\begin{array}{ccccc}
-0.009 & 0.094 & 0.090 & -0.991 & 0.027 \\
0.879 & -0.443 & -0.157 & -0.066 & -0.041 \\
0.452 & 0.888 & -0.005 & 0.0787 & -0.028 \\
0.141 & -0.075 & 0.982 & 0.080 & -0.053 \\
0.056 & 0 & 0.043 & 0.031 & 0.997 
\end{array}
\right),
\end{equation}
and we get the approxmated relation 
\begin{equation}
\left(
\begin{array}{c}
\frac{\Delta \ell_{1}}{\ell_{1}} \\
\frac{\Delta H_{1}}{H_{1}} \\
\frac{\Delta H_{2}}{H_{2}} \\
\frac{\Delta H_{3}}{H_{3}} 
\end{array}
\right) = DV^{\mathrm{T}}\left(
\begin{array}{c}
\frac{\Delta \omega_{b}}{\omega_{b}} \\
\frac{\Delta \omega_{m}}{\omega_{m}} \\
\frac{\Delta h}{h} \\
\frac{\Delta n_{s}}{n_{s}} \\
\frac{\Delta \tau}{\tau} 
\end{array}
\right) \approx D\left(
\begin{array}{c}
-0.991 \frac{\Delta n_{s}}{n_{s}} \\
0.879 \frac{\Delta \omega_{b}}{\omega_{b}} -0.443 \frac{\Delta\omega_{m}}{\omega_{m}} \\
0.452 \frac{\Delta \omega_{b}}{\omega_{b}}+0.888\frac{\Delta\omega_{m}}{\omega_{m}} \\
0.983 \frac{\Delta h}{h} \\
0.997 \frac{\Delta \tau}{\tau} 
\end{array}
\right). \label{append_eq2}
\end{equation}
This tells us that the four standard cosmological parameters, $n_{s}, \omega_{b}, \omega_{m}$ and $h$ can be determined by observing the four variables $\ell_{1}, H_{1}, H_{2}$ and $H_{3}$, however the optical depth $\tau$ cannot be limited sufficiently, which is the direct consequence of the fact that we cannot tightly constrain $\tau$ only through the temperature anisotropy spectrum.

In the next case, we add one physical constant $\Delta\alpha/\alpha$ to the parameter vector;
\begin{equation}
\left(
\begin{array}{c}
\frac{\Delta \ell_{1}}{\ell_{1}} \\
\frac{\Delta H_{1}}{H_{1}} \\
\frac{\Delta H_{2}}{H_{2}} \\
\frac{\Delta H_{3}}{H_{3}} 
\end{array}
\right)=\left(
\begin{array}{cccccc}
0.069 & -0.109 & -0.206 & 0.162 & 0 & -1.442 \\
0.414 & -0.408 & -0.294 & 2.666 & -0.093 & 0.644 \\
-0.647 & -0.045 & 0 & 1.183 & 0 & 1.673 \\
-0.415 & 0.438 & 0 & 0.760 & 0 & 0.967 
\end{array}
\right) \left(
\begin{array}{c}
\frac{\Delta \omega_{b}}{\omega_{b}} \\
\frac{\Delta \omega_{m}}{\omega_{m}} \\
\frac{\Delta h}{h} \\
\frac{\Delta n_{s}}{n_{s}} \\
\frac{\Delta \tau}{\tau} \\
\frac{\Delta \alpha}{\alpha} 
\end{array}
\right).
\end{equation}
From the singular value decomposition and the relevant approximation, we obtain
\begin{align}
D&=\left(
\begin{array}{cccccc}
-0.694 & -1.219 & -0.436 & -0.094 & 0 & 0 \\
2.540 & -1.2123 & 0.172 & 0.045 & 0 & 0 \\
1.970 & 0.810 & -0.162 & -0.239 & 0 & 0 \\
1.182 & 0.541 & -0.354 & 0.246 & 0 & 0
\end{array}
\right) \\
V^{\mathrm{T}} &=\left(
\begin{array}{cccccc}
-0.062 & -0.044 & -0.049 & 0.810 & -0.019 & 0.579 \\
-0.342 & 0.212 & 0.156 & -0.528 & 0.029 & 0.731 \\
0.789 & -0.459 & 0.106 & -0.200 & -0.043 & 0.337 \\
0.500 & 0.861 & 0.048 & 0.069 & -0.032 & 0.025 \\
0.056 & 0 & 0.051 & 0.032 & 0.997 & -0.001 \\
0.062 & 0.029 & -0.978 & -0.141 & 0.052 & 0.125 
\end{array}
\right)
\end{align}
and 
\begin{equation}
\left(
\begin{array}{c}
\frac{\Delta \ell_{1}}{\ell_{1}} \\
\frac{\Delta H_{1}}{H_{1}} \\
\frac{\Delta H_{2}}{H_{2}} \\
\frac{\Delta H_{3}}{H_{3}} 
\end{array}
\right) = DV^{\mathrm{T}}\left(
\begin{array}{c}
\frac{\Delta \omega_{b}}{\omega_{b}} \\
\frac{\Delta \omega_{m}}{\omega_{m}} \\
\frac{\Delta h}{h} \\
\frac{\Delta n_{s}}{n_{s}} \\
\frac{\Delta \tau}{\tau} \\
\frac{\Delta \alpha}{\alpha} 
\end{array}
\right) \approx D\left(
\begin{array}{c}
0.810 \frac{\Delta n_{s}}{n_{s}} +0.579 \frac{\Delta\alpha}{\alpha} \\
-0.342 \frac{\Delta \omega_{b}}{\omega_{b}} -0.528 \frac{\Delta n_{s}}{n_{s}}+0.731 \frac{\Delta \alpha}{\alpha} \\
0.789 \frac{\Delta \omega_{b}}{\omega_{b}}-0.459 \frac{\Delta\omega_{m}}{\omega_{m}} +0.337 \frac{\Delta\alpha}{\alpha} \\
0.500 \frac{\Delta \omega_{b}}{\omega_{b}} + 0.861 \frac{\Delta \omega_{m}}{\omega_{m}} \\
0.997 \frac{\Delta \tau}{\tau} \\
-0.978 \frac{\Delta h}{h} 
\end{array}
\right). \label{append_eq3} 
\end{equation}
The last equation (\ref{append_eq3}) makes a contrast with (\ref{append_eq2}). The newly included parameter $\Delta\alpha/\alpha$ is now determined together with $n_{s}$, $\omega_{b}$ and $\omega_{m}$, on the other hand the Hubble parameter $h$ becomes the unlimited parameter like $\tau$. This is exactly what is expected because in the previous paper \cite{Nakashima:2008cb} we have made sure that, in constraining $\Delta\alpha/\alpha$, the HST prior on the Hubble parameter is very useful, which implied that including $\Delta\alpha/\alpha$ in the analysis weakens the constraint on $h$.

As the third case, which we are most interested in, we add the new parameter $\Delta m_{p}/m_{p}$;
\begin{equation}
\left(
\begin{array}{c}
\frac{\Delta \ell_{1}}{\ell_{1}} \\
\frac{\Delta H_{1}}{H_{1}} \\
\frac{\Delta H_{2}}{H_{2}} \\
\frac{\Delta H_{3}}{H_{3}} 
\end{array}
\right)=\left(
\begin{array}{cccccc}
0.069 & -0.109 & -0.206 & 0.162 & 0 & 0.087 \\
0.414 & -0.408 & -0.294 & 2.666 & -0.093 & 0.127 \\
-0.647 & -0.045 & 0 & 1.183 & 0 & -0.783 \\
-0.415 & 0.438 & 0 & 0.760 & 0 & -0.567 
\end{array}
\right) \left(
\begin{array}{c}
\frac{\Delta \omega_{b}}{\omega_{b}} \\
\frac{\Delta \omega_{m}}{\omega_{m}} \\
\frac{\Delta h}{h} \\
\frac{\Delta n_{s}}{n_{s}} \\
\frac{\Delta \tau}{\tau} \\
\frac{\Delta m_{p}}{m_{p}} 
\end{array}
\right). \label{append_eq4}
\end{equation}
Similar decomposition and approximation lead to
\begin{align}
D&=\left(
\begin{array}{cccccc}
-0.177 & 0.170 & -0.056 & 0.172 & 0 & 0 \\
-2.676 & 0.629 & 0.0305 & -0.017 & 0 & 0 \\
-1.260 & -0.890 & -0.230 & -0.006 & 0 & 0 \\
-0.776 & -0.763 & 0.281 & 0.031 & 0 & 0 
\end{array}
\right) \\
V^{\mathrm{T}}&=\left(
\begin{array}{cccccc}
0.002 & 0.088 &	0.088 & -0.985 & 0.027 & 0.114 \\
0.647 & -0.316 & -0.122 & 0.040 & -0.033 & 0.681 \\
0.301 & 0.936 &	0.0190 & 0.102 & -0.0210 & 0.145 \\
-0.139 & 0.075 & -0.983 & -0.080 & 0.053 & -0.002 \\
0.0380 & 0.003 & 0.046 & 0.033 & 0.997 & 0.019 \\
-0.685 & 0.098 & 0.095 & 0.098 & 0.004 & 0.709 
\end{array}
\right)
\end{align}
and 
\begin{equation}
\left(
\begin{array}{c}
\frac{\Delta \ell_{1}}{\ell_{1}} \\
\frac{\Delta H_{1}}{H_{1}} \\
\frac{\Delta H_{2}}{H_{2}} \\
\frac{\Delta H_{3}}{H_{3}} 
\end{array}
\right) = DV^{\mathrm{T}}\left(
\begin{array}{c}
\frac{\Delta \omega_{b}}{\omega_{b}} \\
\frac{\Delta \omega_{m}}{\omega_{m}} \\
\frac{\Delta h}{h} \\
\frac{\Delta n_{s}}{n_{s}} \\
\frac{\Delta \tau}{\tau} \\
\frac{\Delta m_{p}}{m_{p}} 
\end{array}
\right) \approx D\left(
\begin{array}{c}
-0.985 \frac{\Delta n_{s}}{n_{s}} \\
-0.647 \frac{\Delta \omega_{b}}{\omega_{b}}+0.681 \frac{\Delta m_{p}}{m_{p}} \\
0.936 \frac{\Delta\omega_{m}}{\omega_{m}} \\
-0.983 \frac{\Delta h}{h} \\
0.997 \frac{\Delta \tau}{\tau} \\
-0.685 \frac{\Delta \omega_{b}}{\omega_{b}}+0.709 \frac{\Delta m_{p}}{m_{p}} 
\end{array}
\right). 
\end{equation}
It is found that there are two degeneracy directions and one of them is $-0.685 \Delta \omega_{b}/\omega_{b}+0.709\Delta m_{p}/m_{p}$. This means that $\omega_{b}$ and $m_{p}$ are strongly degenerate, which could be naively expected from just comparing the coefficients of the two parameters $\Delta\omega_{b}/\omega_{b}$ and $\Delta m_{p}/m_{p}$. 

Now, we include the small scale information in the analysis to study how the damping tail is crucial for parameter determination from CMB in the third case. We make use of  $H_{4}$ and $H_{5}$, which is defined in the same way as $H_{2}, H_{3}$, that is,
\begin{equation}
H_{4} \equiv \left( \frac{\Delta T_{\ell_{4}}}{\Delta T_{\ell_{1}}}\right)^{2}, \ \ H_{5} \equiv \left( \frac{\Delta T_{\ell_{5}}}{\Delta T_{\ell_{1}}}\right)^{2}.
\end{equation}
In other words, $H_{4}$ is the ratio of the fourth peak $\ell_{4}$ to the first peak $\ell_{1}$ height and $H_{5}$ is the ratio of the fifth peak $\ell_{5}$ to the first peak $\ell_{1}$. Adding the relations between $\Delta H_{4}/H_{4}, \Delta H_{5}/H_{5}$ and the variations of the cosmological parameters, (\ref{append_eq4}) becomes 
\begin{equation}
\left(
\begin{array}{c}
\frac{\Delta \ell_{1}}{\ell_{1}} \\
\frac{\Delta H_{1}}{H_{1}} \\
\frac{\Delta H_{2}}{H_{2}} \\
\frac{\Delta H_{3}}{H_{3}} \\
\frac{\Delta H_{4}}{H_{4}} \\
\frac{\Delta H_{5}}{H_{5}}
\end{array}
\right)=\left(
\begin{array}{cccccc}
0.069 & -0.109 & -0.206 & 0.162 & 0 & 0.087 \\
0.414 & -0.408 & -0.294 & 2.666 & -0.093 & 0.127 \\
-0.647 & -0.045 & 0 & 1.183 & 0 & -0.783  \\
-0.415 & 0.438 & 0 & 0.760 & 0 & -0.567 \\
-0.478 & 0.315 & 0 & 1.608 & 0 & -0.868 \\
0.665 & 0.493 & 0 & 1.828 & 0 & -0.794 
\end{array}
\right) \left(
\begin{array}{c}
\frac{\Delta \omega_{b}}{\omega_{b}} \\
\frac{\Delta \omega_{m}}{\omega_{m}} \\
\frac{\Delta h}{h} \\
\frac{\Delta n_{s}}{n_{s}} \\
\frac{\Delta \tau}{\tau} \\
\frac{\Delta m_{p}}{m_{p}} 
\end{array}
\right).
\end{equation} 
We can recast this equation through the relevant approxmation in
\begin{equation}
\left(
\begin{array}{c}
\frac{\Delta \ell_{1}}{\ell_{1}} \\
\frac{\Delta H_{1}}{H_{1}} \\
\frac{\Delta H_{2}}{H_{2}} \\
\frac{\Delta H_{3}}{H_{3}} \\
\frac{\Delta H_{4}}{H_{4}} \\
\frac{\Delta H_{5}}{H_{5}}
\end{array}
\right)\approx\left(
\begin{array}{cccccc}
0.1378 & -0.198 & 0.085 & -0.012 & -0.165 & 0.001 \\
2.528 &	-1.024 & 0.343 & -0.042 & 0.023 & -0.001 \\
1.332 &	0.680 & 0.409 &	0.170 &	-0.013 & -0.003 \\
0.897 &	0.648 &	-0.027 & -0.194 & -0.014 & -0.004 \\
1.785 &	0.684 & 0.104 &	-0.0469 & 0.012 & 0.006 \\
2.007 &	-0.046 & -0.789 & 0.069 & -0.013 & -0.001 
\end{array}
\right) \left(
\begin{array}{c}
0.960 \frac{\Delta n_{s}}{n_{s}} \\
-0.617 \frac{\Delta \omega_{b}}{\omega_{b}}-0.658\frac{\Delta m_{p}}{m_{p}} \\
-0.734 \frac{\Delta \omega_{b}}{\omega_{b}}-0.579\frac{\Delta\omega_{m}}{\omega_{m}}  \\
-0.728\frac{\Delta\omega_{m}}{\omega_{m}}-0.582 \frac{\Delta m_{p}}{m_{p}} \\
0.959 \frac{\Delta h}{h} \\
0.994 \frac{\Delta \tau}{\tau} 
\end{array}
\right).
\end{equation} 
All the elements of the last column in $6\times 6$ matrix, which correspond to the coefficients of the parameter $\tau$, are nearly zero, which is almost the same situation occurred in the previous case without $\Delta H_{4}/H_{4}$ and $\Delta H_{5}/H_{5}$. Unlike the previous case, there are no other (nearly) zero columns, which states that, as mentioned in section \ref{sec:conclusion}, the strong degeneracy between $\omega_{b}$ and $m_{p}$ can be resolved by observing not only $\ell_{1}, H_{1}, H_{2}, H_{3}$ but also $H_{4}$, $H_{5}$ which possess the damping tail information.

\section*{Appendix B}
In this appendix, we briefly report a forecast for constraints from the future CMB survey, in particular, the PLANCK survey \cite{Planck:2006uk}. The parameters for the instrumental design for PLANCK survey are shown in TABLE \ref{hyou2} and mock data are generated by the publicly available FuturCMB code \cite{Perotto:2006rj}. The fiducial parameter values are the same as those adopted in section \ref{sec:effect} and we perform the MCMC analysis in the same way as in section \ref{sec:constraint}. Throughout this appendix, we assume the model between time variations of physical constants introduced in section \ref{sec:dilaton}. 

\begin{table}
\caption{Assumed instrumental parameters for PLANCK survey. $f_{sky}$ is the fraction of the observed sky, $\theta_{\mathrm{FWHM}}$ is Gaussian beam width at FWHM, $\sigma_{T}$ and $\sigma_{P}$ are temperature and polarization noise, respectively.}
\label{hyou2}
\begin{center}
\begin{tabular}{|c|c|c|c|c|}
\hline 
$f_{sky}$ & bands[GHz] & $\theta_{\mathrm{FWHM}}[arcmin]$ & $\sigma_{T}$[$\mu$K] & $\sigma_{P}$[$\mu$K] \\
\hline 
0.65 & 100 & 9.5 & 6.8 & 10.9 \\
     & 143 & 7.1 & 6.0 & 11.4 \\
     & 217 & 5.0 & 13.1 & 26.7 \\   
\hline
\end{tabular}
\end{center}
\end{table}

In TABLE \ref{hyou3}, we show the forecast for the constraints on the cosmological parameters as the mean values, 68$\%$ and 95$\%$ confidence intervals. As is commented in the concluding section and appendix A, future CMB survey can constrain time variation of the physical constants more tightly and the limit on $\Delta \alpha/\alpha$ or $\Delta \mu/\mu$ from PLANCK experiment is expected to be about one order tighter than the constraint from the WMAP 5-year data and comparable to the constraint from BBN \cite{Ichikawa:2002bt}.

\begin{table}
\caption{MCMC results on the mean values, 68$\%$ and 95$\%$ confidence intervals of cosmological parameters.}
\label{hyou3}
\begin{center}
\begin{tabular}{|l|c|}
\hline 
  & varying $\alpha, m_{e}$ and $m_{p}$ \\
\hline \hline
$100\Omega_{B}h^{2}$ & $2.274^{+0.025 +0.050}_{-0.025 -0.049}$ \\ \hline
$\Omega_{DM}h^{2}$ & $0.111^{+0.002 +0.003}_{-0.002 -0.003}$ \\ \hline
$\tau$ & $0.087^{+0.002 +0.009}_{-0.003 -0.009}$ \\ \hline
$n_{s}$ & $0.963^{+0.009 +0.017}_{-0.009 -0.018}$  \\ \hline 
$\log(10^{10} A_{s})$ & $3.065^{+0.011 +0.023}_{-0.011 -0.023}$ \\ \hline
$H_{0}$ & $71.9^{+1.0 +2.1}_{-1.1 -2.0}$ \\ \hline
$\Delta\alpha/\alpha$ & $0.0000021^{+0.0002 +0.0003}_{-0.0002 -0.0003}$ \\ \hline
$\Delta \mu/\mu$ & $0.0003^{+0.016 +0.030}_{-0.015 -0.031}$ \\ \hline
\end{tabular}
\end{center}
\end{table}

\end{document}